\documentclass[prb,preprint]{revtex4}
\usepackage{amsmath}
\usepackage{graphicx}

\begin{document}

\title{Genesis and the tipping pencil; why the Universe is flat}

\author{Ronald J. Adler}
\email{adler@relgyro.stanford.edu}
\affiliation{Hansen Experimental Physics Laboratory, Gravity Probe B Mission, Stanford University, Stanford, California 94309}
\affiliation{Department of Physics and Astronomy, San Francisco State University, San Francisco, California 94132}


\begin{abstract}
In a room with five cosmologists there there may easily be ten theories of cosmogenesis. Cosmogenesis is a popular topic for speculation because it is philosophically deep and because such speculations are unlikely to be proven wrong in the near future. The scenario we present here was intended mainly as a pedagogical illustration or toy model, but it turns out to possibly have a more serious and interesting result - a rationale for the spatial flatness of the Universe. Our basic assumptions are that the cosmological scale factor obeys the standard Friedmann equation of general relativistic cosmology and that the equation is dominated by a cosmological constant term and a curvature term; the dynamics of the Universe is then (approximately) the same as that of a tipping pencil. The scale factor cannot remain at an unstable initial value of zero and must increase (i.e. the Universe must come into existence) according to the uncertainty principle, that is due to quantum fluctuations; in other words we propose in a precise but limited context an answer to Heidegger's famous question ``Why is there something rather than nothing." The mechanism is the same as that whereby an idealized pencil balanced on its point cannot remain so and must tip over. If it is moreover assumed that the Universe expands at the minimum asymptotic rate consistent with the uncertainty principle then the result is spatial flatness. 

\end{abstract}

\maketitle

\section{Introduction}

Cosmology has been a remarkably successful field in the last half century. \cite{1} Due to our understanding of high energy particle physics embodied in the standard model of quarks and leptons we have considerable confidence in our understanding of the Universe as far back as a microsecond or less and may even make reasonable theoretical speculations back to the inflation era at roughly 
$10^{-36}$ \rm{s}. \cite{2} For times before that we are free to speculate with varying degrees of plausibility; there has been a great amount of such speculation, although it is often not clearly labeled as speculation. \cite{3}

One remarkable observational fact concerning the $\Lambda$CDM model is that the current Universe is quite close to being flat; that is $\Omega = 0$ to about a percent, which inflationary theorists ascribe to the large and rapid increase of the scale factor.  \cite{1,4} Our main result is a plausible alternative explanation for the flatness.  

Our basic assumption in this note is that the scale factor of the Universe at $t = 0$ is described by the usual Friedmann equation, which is dominated by a cosmological constant term and a curvature term. \cite{5} Then the dynamics of the scale factor is approximately the same as that of the idealized tipping pencil shown in fig. 1, with an infinitely sharp point. The pencil thus serves as an amusing analog computer. 

A well-known homework problem in quantum mechanics is to calculate the the maximum time that such a pencil can be balanced before it tips over, subject to the uncertainty principle. \cite{6} The same calculation can be applied to the scale factor of the Universe to see why it cannot remain at zero and how rapidly it must increase. Curiously it turns out that the analogy leads to a spatially flat Universe. 

In sec. 2 we review the analysis of the tipping pencil, treated semi-classically but subject to the uncertainty principle. In section 3 we analyze the evolution of the Universe as an unstable system analogous to the pencil, with its evolution initiated by a quantum fluctuation according to the uncertainty principle, which leads to a spatially flat Universe. Within the context of cosmology and precise assumptions about the contents of the Universe we therefore offer an answer to Heidegger's famous question ``Why is there something rather than nothing."\cite{6.5}

Almost needless to say our cosmogenesis scenario is quite speculative. We call it a scenario or toy model since it is not based on a compete theory, but it does give a plausible explanation of spatial flatness. We emphasize that the model is not a quantum cosmology but only uses a quantum fluctuation to get things started. \cite{7}
\section{Parable of the tipping pencil}

We first consider the simplified pencil resting on its tip in fig.1. It is easy to obtain its equation of motion for small angles, $\theta < 1$,
\begin{equation}
\label{1}
\ddot{\theta}=\omega^2\sin\theta\cong\omega^2\theta.
\end{equation}
Here $\omega$ is the frequency the pencil would have if suspended by its tip as a pendulum, 
 $\omega=\sqrt{g/l}$. A typical value is about $\omega\cong10 \rm{rad/s}$. The solution of Eq.~(\ref{1}) in terms of the initial angular position $\theta_{0}$ and angular velocity $\dot\theta_{0}$ is  
\begin{equation}
\label{2}
\theta= \theta_{0}\cosh(\omega t)+ (\dot\theta_{0}/\omega)\sinh(\omega t).                     
\end{equation}
Classically the pencil could be balanced forever, $\theta = 0$, by choosing the initial conditions 
$\theta_{0}=0$ and $\dot\theta_{0}=0$, although of course this solution is unstable. 

The initial conditions $\theta_\text{0}=0$ and $\dot\theta_{0}=0$ are not consistent with the uncertainty principle. If we denote the mass by $m$ and the effective length by $\l$ then the initial uncertainties in position and momentum for the top of the pencil are 	
\begin{subequations}
\begin{align} 
\label{3a}
\Delta x\approx \theta_{0}\l,\\
\label{3b}
\Delta p\approx{m}\dot\theta_{0}\l,
\end{align}
\end{subequations}
so the uncertainty principle demands
\begin{equation}
\label{4}
\theta_{0}\dot \theta_{0} > \hbar/m\l^2.                      
\end{equation}
We obtain the maximum time one can balance the pencil, consistent with the uncertainty principle, by taking  
 \begin{equation}
\label{5}
 \dot \theta_{0}=\hbar/m\l^2\theta_{0},                      
 \end{equation} 
and minimizing $\theta$ with respect to $\theta_{0}$ in Eq.~(\ref{2}). This gives
\begin{equation}
\label{6}
(ml^2\omega/\hbar) \theta_{0}^2=\tanh(\omega{t}),                      
\end{equation} 
The tanh function rapidly approaches 1 for $\omega{t}>1$, so to maximize the time of fall we choose 
$ \theta_\text{0}^2=\hbar/(ml^2\omega)$ and thus obtain the initial values and solution to be  
\begin{subequations}
\begin{align} 
\label{7a}
\theta_{0}=\dot\theta_{0}/\omega=\sqrt{\hbar/(ml^2\omega)},\\
\label{7b}
\theta=\sqrt{\hbar/(ml^2\omega)}[\cosh(\omega{t})+\sinh(\omega{t})]=\sqrt{\hbar/(ml^2\omega)}\exp(\omega{t})).
\end{align}
\end{subequations}
Of course this solution is only valid for $\theta < 1$.

A first integral of Eq.~(\ref{1}) is 
\begin{equation}
\label{8}
\dot \theta^2 / \theta^2 = \omega^2 - k / \theta^2,                  
 \end{equation}
and the constant of integration $k$ may be expressed in terms of the initial conditions as 
\begin{equation}
\label{9}
k= \omega^2 \theta_{0}^2- \dot\theta_{0}^2.                  
 \end{equation}
The effective potential is thus an upside-down harmonic oscillator, $V=-\omega^2\theta^2$. Eq.~(\ref{8})
is the same as the cosmological equation we will discuss in the next section, except that the restriction 
$\theta < 1$ will not apply to the cosmological problem. 

It is amusing to calculate a typical maximum value for the time it takes the pencil to fall, which is surprisingly short. From Eq.~(\ref{7b}) the time to fall to  $\theta$ is 	
\begin{equation}
\label{10}
t=(1/ \omega)[\ln(\theta) +(1/2)\ln(ml^2\omega/\hbar)].              
 \end{equation}	 						 
For a typical pencil rough values are $m\approx10^{-2} \rm{kg}$  and $l\approx10^{-1} \rm{m}$  so that  $\omega\approx10 \rm{rad} / \rm{s}$ and 
\begin{equation}
\label{11}
t=(1/10)[\ln(\theta) +36)] \rm{s}\approx3.6 \rm{s}.                       
 \end{equation}	 							
The time is very insensitive to the choice of  $\theta$ as the final angle. One can easily balance a pencil for about a second, so this quantum-imposed maximum is not much longer. Finally, note that in one second $\omega t\approx 10$  and $\tanh(\omega t)\approx 1- 4 \times 10^{-9}$, so taking $\tanh(\omega t)=1$  is quite well justified even for relatively short times.						
\section{ Cosmogenesis}

From the amusing but mundane problem of the tipping pencil we proceed to the deeper and more interesting problem of the origin of the Universe and its spatial geometry. From the standard starting point of the Einstein equations applied to the Freidmann-Robertson-Walker metric we obtain the Friedmann equation for the cosmological scale factor ${a}$, which is the fundamental equation of cosmology,
\begin{equation}
\label{12}
\dot a^2 / a^2 = (8\pi{G}\rho/3)+(\Lambda/3) - k / a^2.                  
\end{equation}
Here $\rho$  is the density of any fluid present in the early Universe, not including dark energy,   $\Lambda$ is the cosmological constant at early time (presumably very much larger than at the present value), and $k$  is the spatial curvature. \cite{5} For a hyperspherical universe the curvature is positive,  for a pseudo-hyperspherical universe the curvature is negative, and for a flat universe it is zero.

For the earliest times it is plausible to assume that the Universe contains no fluid density and is dominated by the cosmological constant (equivalent to dark energy) and the curvature term of unknown sign and magnitude. This assumption is essentially the same as made in a typical inflation theory, except that a slowly varying field plays the role of the cosmological constant. \cite{1} Thus we deal with 
\begin{equation}
\label{13}
\dot a^2 / a^2 =(\Lambda/3) - k / a^2.                  
\end{equation}						
We recognize this as the same equation as the first integral of the tipping pencil Eq.~(\ref{8}) with the correspondence $\theta\equiv a$  and $\omega\equiv\sqrt(\Lambda/3)$ , so we can immediately write the solution in terms of the initial values $a_{0}$  and $\dot a_{0}$ as the analog of Eq.~(\ref{2})
\begin{equation}
\label{14}
a= a_{0}\cosh(\omega t)+ (\dot a_{0}/\omega)\sinh(\omega t)                      
\end{equation}																		The curvature in terms of the initial values is 
\begin{equation}
\label{15}
k= \omega^2 a_{0}^2- \dot a_{0}^2.                  
 \end{equation}
The cosh and sinh solutions in Eq.~(\ref{14}) are of course well-known and correspond to positive and negative values of the curvature.  

The amusing physics now enters if we ask why the Universe should exist; that is why should the Universe not remain in unstable equilibrium at $a=0$ , rather than have $a > 0$  which we observe so readily? A reasonable answer is that an uncertainty principle should prohibit  $a_{0}=\dot a_{0}=0$ so that a quantum fluctuation pushes the Universe out of its unstable equilibrium condition, precisely as with the pencil in section 2.  As with the uncertainty principle Eq.~(\ref{4}) for the pencil such an uncertainty principle should take the form 
\begin{equation}
\label{16}
a_{0}\dot a_{0} > \hbar/m\l^2.                      
\end{equation}
where $m$  and $l$  are some characteristic mass and length parameters of the nascent Universe, perhaps the Planck mass and length for example. Thus we ask that the Universe have a minimum scale factor at asymptotically large times consistent with the uncertainty principle and obtain, as in Eq.~(\ref{7b}), 
\begin{equation}
\label{17}
a=\sqrt{\hbar/(ml^2\omega)}\exp(\omega {t}).
 \end{equation} 							
Therefore the scale factor is prevented from being zero by the nonzero value of $\hbar$  and the Universe is forced into existence by the uncertainty principle. Moreover the demand of minimum expansion leads to zero spatial curvature, which is a rather interesting result. 
	
We could ask what reasonable values the parameters $m$  and $l$  might have and what the coefficient in Eq.~(\ref{17}) would then be. However for $k=0$  the Friedmann Eq.~(\ref{13}) is scale invariant so the scale factor is arbitrary up to a constant multiplier. The coefficient in Eq.~(\ref{17}) is thus not observable and the values of  $m$ and  $l$ are irrelevant. 

In summary, the uncertainty principle Eq.~(\ref{16}) serves to destroy the unstable equilibrium and start the expansion of the Universe, but any further effects, other than spatial flatness, are thereby erased. 
\section{Summary}

We have presented a scenario for cosmogenesis in which the nascent Universe is dominated by a cosmological constant, or dark energy, and curvature. The Universe is prevented from remaining in unstable equilibrium with a scale factor of $a=0$ by the uncertainty principle. If the subsequent expansion is taken to be minimal then the curvature must be zero. That is we may view the flatness as the result of a ``laziness principle." The scenario is not a quantum cosmology although a quantum effect serves to begin the expansion. There is thus no difficulty of interpretation common to quantum cosmologies. 
\section{Acknowledgements}
	
This cosmogenesis scenario was developed as an example for a course on the Big Bang at San Francisco State University.  I thank the students in the course for their comments. Robert Wagoner,  James Bjorken and Pisin Chen later provided interesting and valuable discussions and criticisms.

\section*{Figure captions}
\begin{figure}[h!]
\includegraphics[width=5cm]{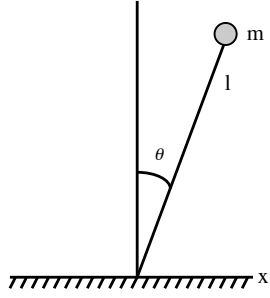}
\caption{The schematic pencil with mass $m$ and length $l$ tips at angle $\theta$.}
\end{figure}


\begin{thebibliography}{99}
\bibitem{1} W. L. Freedman and E. W. Kolb,  {``Cosmology,"} R. J. Adler,   {``Gravity,"}  A. Dar,  {``The new astonomy,"} in \textit{The New Physics for the Twenty-first Century}, edited by Gordon Fraser (Cambridge University Press, Cambridge UK, 2006). 
\bibitem{2} C. Quigg, {``Particles and the standard model,"} in \textit{The New Physics for the Twenty-first Century}, edited by Gordon Fraser (Cambridge University Press, Cambridge UK, 2006).
\bibitem{3} A. Liddle, \textit{An Introduction to Modern Cosmology}, (John Wiley, West Sussex UK, 2003), chapter 14. S. W. Hawking and G. F. R. Ellis, \textit{The Large Scale Structure of Space-Time}, (Cambridge University Press, Cambridge UK, 1973), chapter 10.
\bibitem{4} R. J. Adler and J. M. Overduin, Gen. Relativ. Gravit. \textbf{37}, 1491(2005). 
R. P. Feynman, \textit{Lectures on Gravitation}, notes taken by F. Morinigo and W. G. Wagner (California Institute of Technology, Pasadena CA, 1971), Lecture 3.
\bibitem{5} See chapters 3 and 4 of Liddle in reference 3.
\bibitem{6} R. H. Dicke and J. P. Wittke, \textit{Introduction to Quantum Mechanics 1st edn.} (Addison-Wesley, Reading MA, 1960), chapter 2; an objection to the assumption of an infinitely sharp point is made by D. Easton, Eur. J. Phys. \textbf{28}, 1097 (2007) We find the objection to be  ``not even wrong." 
\bibitem{6.5} There appears to be some controversy as to the origin and precise statement of the question. See \url{http://plato.stanford.edu/entries/nothingness/}; M. Heidegger, \textit{Introduction to Metaphysics}, trans. R. Manheim (Yale University Press, New Haven CT, 1959); M. Heidegger, \textit{Being and Time}, trans. J. McQuarrie and E. Robinson (Harper and Row, New York NY, 1962).
\bibitem{7} For an overview of quantum cosmology, L. Z. Fang and Z. C. Wu, Int. J. Mod. Phys. A, \textbf{1}, 887 (1986); a later survey, D. L. Wiltshire, \url{arxiv:gr-qc/0101003v2}.
\end{thebibliography}
\end{document}